\documentclass{elsarticle}

\begin{document}

\title{\textbf{SOM-based DDoS Defense Mechanism using SDN for the Internet of Things}}

\author[1]{Yunfei Meng}
\author[1]{Zhiqiu Huang}
\author[1]{Senzhang Wang}
\author[1]{Guohua Shen}
\author[2]{Changbo Ke}

\address[1]{College of Computer Science and Technology, Nanjing University of Aeronautics and Astronautics, Nanjing 211106, China}
\address[2]{School of Computer Science and Technology, Nanjing University of Posts and Telecommunications, Nanjing 210023, China}

\begin{abstract}
To effectively tackle the security threats towards the Internet of things, we propose a SOM-based DDoS defense mechanism using software-defined networking (SDN) in this paper. The main idea of the mechanism is to deploy a SDN-based gateway to protect the device services in the Internet of things. The gateway provides DDoS defense mechanism based on SOM neural network. By means of SOM-based DDoS defense mechanism, the gateway can effectively identify the malicious sensing devices in the IoT, and automatically block those malicious devices after detecting them, so that it can effectively enforce the security and robustness of the system when it is under DDoS attacks. In order to validate the feasibility and effectiveness of the mechanism, we leverage POX controller and Mininet emulator to implement an experimental system, and further implement the aforementioned security enforcement mechanisms with Python. The final experimental results illustrate that the mechanism is truly effective under the different test scenarios.
\end{abstract}


\maketitle

\section{Introduction}

\paragraph{}
We summarize the security threats of the Internet of things as the follows: (1) There are a large number of different types of sensing devices in the Internet of things, such as temperature sensors, infrared sensors or cameras. Normally, these sensing devices have very limited computing abilities, and can only use narrow-band communication protocols (such as Bluetooth, RFID, WiFi or ZigBee) to send their collected data to the edge servers or intelligent terminals near with sensing devices \cite{iot1}. We always call these resource-constrained sensing devices as \emph{weak} sensing devices \cite{iot1}. Due to their constrained computing abilities, it is nearly impossible to deploy complex security authentication mechanisms in weak sensing devices, such as private key authentication, verification code authentication or access control mechanisms. Therefore, how to effectively identify and manage massive weak sensing devices is a major problem for Internet of things now. (2) Sensing devices of the Internet of things are often distributed in different geographical spaces and often unattended, so they are vulnerable to direct physical attacks and different types of cyber attacks \cite{iot3}. If an attacker can control some sensing devices in the Internet of things, he can use these devices to launch large-scale network attacks (such as DDoS attacks) and then paralyze the entire Internet of things system \cite{iot3}\cite{iot4}. (3) Large amount of personal privacy information, such as physiological information, location information, video and audio information, will be generated by the user's body area sensing devices or portable mobile terminals \cite{SZWang}. These privacy information is often stored in the device services closely related to the sensing device. If the edge domain deployed these device services lacks effective security protection mechanism, these privacy information might be illegally accessed or utilized by malicious users, which might cause a potential huge threat to the personal security of users \cite{iot2}. In addition to the above three-aspect security threats, the edge domain of the Internet of things also faces other types of security threats, such as the lack of security policy caused by virtual machine (VM) migration or data leakage caused by container technology \cite{iot1}\cite{iot2}.

\paragraph{}
In order to effectively enforce the security of the Internet of things, we propose a SOM-based DDoS defense mechanism using software-defined networking (SDN)\cite{SDN1}. The main idea of the mechanism is that the security gateway monitors all the sensing devices that access the edge domain, and then analyzes the access behavior of the source devices by using SOM neural network classification algorithm. If it is found that the access behavior of a device constitutes a DDoS attack, the security gateway will add the MAC address of the device to the device blacklist and automatically block it in the future. In this case, the security gateway can automatically identify such attacks whether the attacker leverages untrusted devices or trusted devices, and then effectively enhance the security of the entire Internet of things.

\paragraph{}
The remainder of the paper is structured as follows. Section 2 is a brief introduction of self-organized maps (SOM). Section 3 is the main body of this paper, which presents the mechanism in detail. Section 4 implements the mechanism and evaluates its effectiveness. Section 5 reviews some related works and compares them with our mechanism. Finally, Section 6 concludes this paper.

\section{Self-organized Maps}

\paragraph{}
Self-organized maps (SOM) is an unsupervised neural network learning method proposed by Kohonen in 1982 \cite{SOM}. It imitates the processing method of human brain neurons to process information, carries out self training, and automatically clusters the input eigenvectors. The self-organizing mapping network is divided into two layers, input space and mapping grid, as shown in Figure 2. Each input neuron in the input space is usually an n-dimensional eigenvector. Each input neuron has its corresponding mapping neuron in the mapping grid space. Each mapping neuron has an m-dimensional synaptic weights, and $n=m$, that is to say, the weight vector and the input eigenvector must be equal dimensional.

\begin{figure}[htbp]
\centering
\scalebox{0.42}{\includegraphics{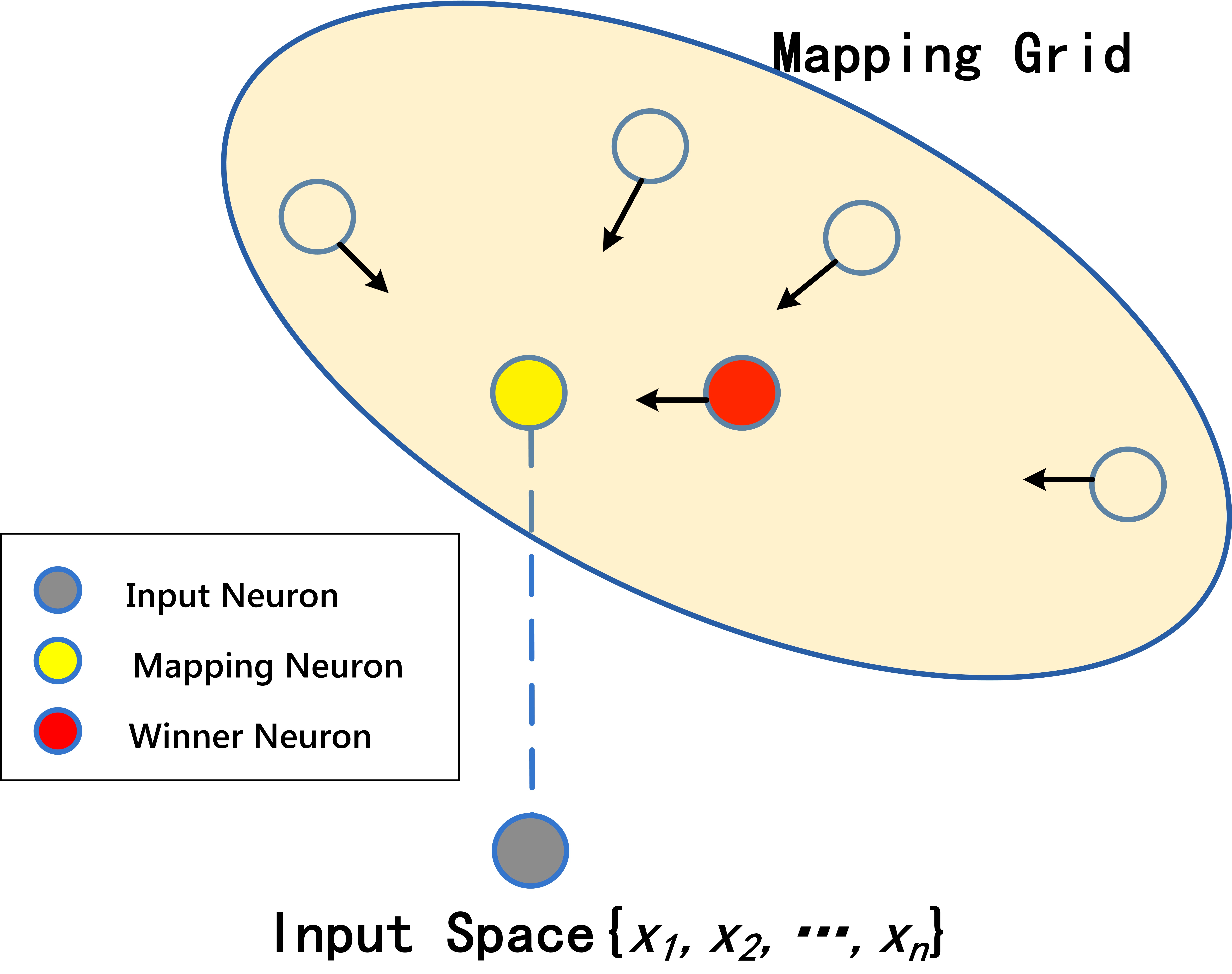}}
\caption{The self-organized maps. }
\end{figure}

\paragraph{}
The self-organizing mapping network uses the competitive learning method to train the grid model, and automatically classifies the input eigenvectors by calculating the similarity between different mapping neurons in the mapping grid. The whole training steps are as follows:

(1) Initialize the grid. Usually, the eigenvector of the input neuron is taken as the initial weight vector of the mapped neuron.

(2) Input a neuron $I$ in the input space, its eigenvector is defined as: $I(x)$=$(x_1$, $x_2$,...,$x_n)$.

(3) In the mapping grid, the mapping neuron of neuron $I$ is defined as: $M$=$(x_1$, $x_2$,...,$x_n)$, then calculate the Euclidean distance or other distance between neuron $M$ and other neurons in the mapping grid. If the weight vector of the mapped neuron equals $W$=$(w_1$, $w_2$,...,$w_n)$, the distance between the mapped neuron and neuron $M$ is calculated as follows:

\begin{equation}
d=\sum_{i=1}^{n}||x_{i}-w_{i}|| \\
\end{equation}

(4) After the calculation, select the mapping neuron $K$ with the minimum distance from neuron $M$ as the winner neuron, that is:
\begin{equation}
d_k=min(\sum_{i=1}^{n}||x_{i}-w_{i}||) \\
\end{equation}

(5) If the weight vector of the winning neuron $K$ equals $W_k$=$(wk_1$, $wk_2$,...,$wk_n)$, then use the value of $W_k$ to adjust the weight vector of other mapping neurons $J$, and the calculation method is as follows:
\begin{equation}
wj_{i}(t+1)=wj_{i}(t)+\eta\cdot\theta(J,K)\cdot(x_{i}(t)-wk_{i}(t)) \\
\end{equation}

\paragraph{}
Where $t$ represents the current time, $0\leq\eta\leq 1$ represents an error coefficient, and $\theta(J,K)$ is a neighborhood function, representing the relative position relationship between neuron $J$ and the winner neuron $K$.

(6) Repeat the step (2) to (5) until the weight vector of all mapped neurons no longer changes significantly, and output the final classification result.

\section{SOM-based DDoS defense mechanism}

\paragraph{}
In order to use SOM neural network for feature classification, we regard the MAC address of each device in the Internet of things as an input neuron $I(x)$. When the device accesses the Internet of things, the OVS will send the packet-in event of the device to the SDN controller for processing. After the controller receives the event, the feature extraction module in the controller will automatically extract the network features set related to the device from the event as the feature vector $(x_1, x_2,...,x_n)$ of the input neuron $I(x)$. At present, we only consider two characteristic values related to DDoS attacks, namely packet growth rate ($X_1$) and packet total number ($X_2$) in window time.

1) The calculation method of packet growth rate ($X_1$) in window time is as follows:
\begin{equation}
X_1=\frac{P(t_i+\triangle t)-P(t_i)}{\triangle t} \\
\end{equation}

Where: $\triangle t$ is the window time, $P(t_i)$ indicates the number of received packets sent from a device at $t_i$ time. The meaning of $X_1$ is that when a device launches a DDoS attack on the Internet of things, compared with other normal devices in a fixed window time, the packet growth rate of the device will increase abnormally and rapidly. Therefore, $X_1$ is a very important characteristic value for us to identify DDoS attacks.

2) The calculation method of the total number of packages $X_2$ is as folows:
\begin{equation}
X_2=\sum_{i=0}^{n}P(\triangle t_i) \\
\end{equation}

Where: $P(\triangle t_i)$ indicates the total number of received packets sent from a device in the $n_{th}$ window time. The meaning of $X_2$ is that when a device launches a DDoS attack on the Internet of things, the total number of packets sent from the device will increase abnormally compared with other normal devices.

\paragraph{}
Next, the SOM neutral network algorithm will automatically read the set of eigenvalues $\{X_1$, $X_2\}$ of all access devices in every window time ($\triangle t$), and then classify and calculate these eigenvalues. In this way, the MAC address of each device corresponds to an input neuron $I(x_i)$, which is expressed as: $I(x_i)$=$\{xi_1$, $xi_2\}$ in the input space. Accordingly, the mapping neuron $M$ in the mapping grid can be expressed as:  $M(x_i)$=$\{xi_1$, $xi_2\}$. According to the calculation (1) and (2), SOM algorithm can find a winner neuron $K(x_i)$ with the highest similarity with the mapping neuron $M(x_i)$, and then use the calculation (3) to update other mapping neurons in the mapping grid to the winner neuron along the gradient direction. In this way, after $N$ times of iterative calculation, SOM neural network algorithm can divide all input neurons into $m$ groups with similar patterns, so we only need to select the group with the largest eigenvalue $X_2$ as the DDoS suspicious group. Since each input neuron $I(x_i)$ in our mechanism corresponds to a unique MAC address of the device, the selected suspicious packet can be transformed into a set $D$=$\{mac_1$, $mac_2$,..., $mac_n\}$. Finally, according to the set $D$, the flow table operation module will send block rules to the OVS so that any suspicious device in set $D$ will be automatically blocked by OVS in the future, thus the mechanism can effectively defend against the DDoS attacks launched by sensing layer.

\section{Implementation and Evaluations}

\paragraph{}
The purpose of this experiment is to verify and evaluate: (1) whether the defense mechanism proposed in this paper can effectively detect DDoS attack. (2) After successfully detecting the DDoS attack, whether the defense mechanism proposed in this paper can make effective dynamic defense, and then enhance the security of the system.

\paragraph{}
First, we validate whether the defense mechanism can effectively detect DDoS attack sources. Here we use the $Ping flood$ instruction in Mininet to simulate the DDoS attacks, i.e., hacker attacks $VM_3$ and physican2 attacks $VM_2$. At 85 seconds after launching the attacks, we start SOM-based DDoS defense mechanism, and then get the real-time monitoring results of the mechanism. According to the evaluation method proposed in reference [Braga], we evaluate the effectiveness of the mechanism. The specific calculation method is as follows:
\begin{equation}
DR=\frac{TP}{TP+FN} \\
\end{equation}

Where: $DR$ (detection rate) indicates the accuracy of detection, $TP$ (true positives) indicates the number of correctly identified malicious attack sources, and $FN$ (false negatives) indicates the number of improperly identified malicious attack sources.

\begin{equation}
FA=\frac{FP}{FP+TN} \\
\end{equation}

Where: $FA$ (false alarm) indicates the false alarm rate, $FP$ (false positives) indicates the number of legitimate sources that are misreported as attack sources, and $TN$ (true negatives) indicates the number of recognized legitimate sources.

\paragraph{}
After multi-round iterations, the defense mechanism has successfully detected 16 source devices in the virtual network, and the number of malicious source devices identified to launch DDoS attacks in the 16 source devices is 2 ($TP$=2), that is, hacker and physician2. The number of malicious attack sources that are not correctly identified is 0 ($FN$=0), the number of legitimate sources that are misreported is 0 ($FP$=0), the number of legitimate source devices that are detected is 14 ($TN$=14), the detection accuracy of the mechanism is 100$\%$ ($DR$=100), and the misinformation rate is 0$\%$ ($FA$=0). Therefore, the experimental results show that our defense mechanism can effectively detect DDoS attack sources.

\paragraph{}
After that, we further validate whether the mechanism can make effective dynamic defense after detecting DDoS attack sources. Here we use $Perfmon$ performance monitoring toolkit to record the network load and CPU load of SDN controller in real time. If the network load and CPU load of SDN controller are significantly reduced after the defense mechanism is started, it means that the mechanism can effectively defend against attacks initiated by DDoS attack sources, otherwise, it means that the mechanism can only effectively detect DDoS attacks, but there is no way to effectively defend against such attacks. Figure 6 shows the real-time performance data of SDN controller recorded by $Perfmon$ in the whole experiment process. The red solid line in the figure represents the network load data of SDN controller, the blue solid line represents the CPU workload of SDN controller, the black dotted line represents the time when DDoS attack starts, and the black solid line represents the time when SOM-based DDoS defense mechanism starts. It can be seen from the figure that in the 120th second of the experiment, we use the hacker and physician2 devices in the virtual network to launch a simulated DDoS attack on the network. This kind of network attack causes a great network load to the SDN controller, that is, the red curve rises rapidly after 120s, and it always fluctuates in the high position. In the 205th second of the experiment, we started the attack defense mechanism. We found that after the defense mechanism was started, the CPU workload (blue line) of the SDN controller fluctuated intermittently and rose, while the network workload (red line) dropped rapidly and returned to normal quickly in the 210th second of the experiment. Here, the CPU workload fluctuates intermittently because our defense mechanism is preset to execute intermittently according to a fixed window time ($\triangle t$). Therefore, the CPU workload of the controller also fluctuates intermittently. The rapid decline of network load fully shows that the defense mechanism has successfully detected the DDoS attack sources, and made effective dynamic defense for these two DDoS attack sources, that is, a large number of packets sent from the attack sources have been directly blocked by the underlying OVS, and will not be forwarded to the SDN controller for analysis and processing, so that the network load of the SDN controller appears a rapid decrease. In conclusion, the experimental results in Figure 5 show that the DDoS defense mechanism proposed in this paper can effectively detect the DDoS attack sources, and can implement effective dynamic defense after successfully detecting the attack sources so that enhance the security of the Internet of things.

\begin{figure}[htbp]
\centering
\scalebox{0.40}{\includegraphics{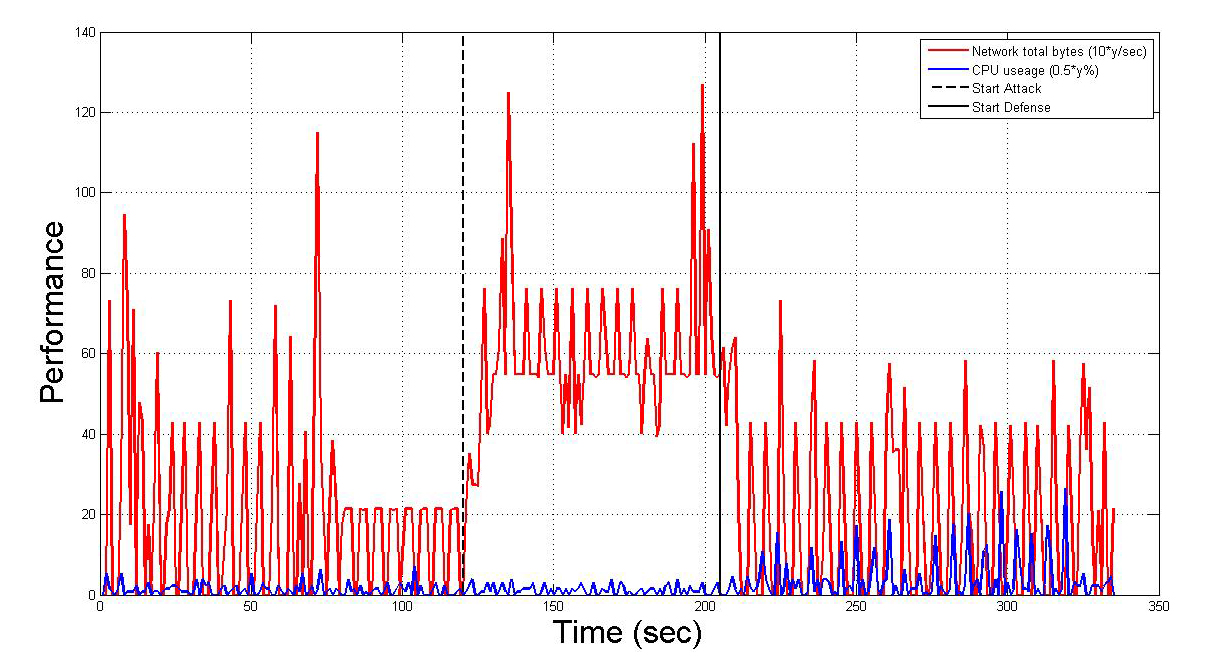}}
\caption{The performance monitoring of SDN controller. }
\end{figure}

\section{Related Work}

\paragraph{}
In this section, we discuss some research works concerning how to implement security enforcement mechanisms using software-defined networking, and compare these proposals with our framework.

\paragraph{}
Hu et al. \cite{SDN9} proposed a comprehensive framework, Flowguard, to facilitate accurate detection as well as flexible resolution of firewall policy violations in dynamic OpenFlow networks. In addition, authors implemented a prototype using Floodlight. The experimental results show that Flowguard has the manageable performance overhead to enable realtime monitoring network. Similarly, Porras et al. \cite{SDN5} proposed a security enforcement controller, FortNOX, which is an extension on NOX controller. FortNOX is designed to enable a network flow to be blocked (or allowed) by security applications. They also proposed a conflict resolving mechanism used in case of appearing policy conflicts. Exactly, we are inspired by the ideas of Flowguard and FortNOX in some sense, we also design the relevant policy resolving mechanism in our framework, i.e., the information flow rules of administrator ($Role=A$) can override those rules of users ($Role=U$). Moreover, we design all entries in OVS can be automatically updated per $t$ minutes, which can also be used to resolve the policy conflicts.

\paragraph{}
Suh et al. \cite{SDN7} leveraged POX controller to implement a firewall application. Each firewall rule can be defined by 6 actions and 12 conditions, and the final experimental results illustrate the firewall is effective. But this mechanism requires network operators to know the details of underlying network, and input the firewall rules into the controller manually. While in our framework, all of information flow rules of IFM are converted from SRM automatically, service providers just need to know which service could be released to which consumer or which thing, other details of underlying network can be created from system models automatically. Therefore, any normal user can leverage our framework to rapidly define their security policies.

\paragraph{}
Koerner et al. \cite{SDN6} proposed a MAC-based VLAN tagging mechanism using SDN. The virtual local area network (VLAN) has been widely used in enterprise networks where the security policy is always defined by VLAN address. But some mobile laptop-based workstations often change their locations, which will leads to the frequent changing of its VLAN address and incur security policy conflicts. To address this problem, authors leverage Floodlight controller to map the MAC address of laptop into its corresponding VLAN address in network. Since MAC address is static, thus it can guarantees the laptops can access the network successfully in different locations. In our framework, the controller use information flow rule to recognize an authorized user, i.e., the pair $\langle$ $Src$, $Dst$ $\rangle$. Here $Src$ is MAC address of service consumer, $Dst$ is VLAN address of VM, but we don't need to convert MAC address into a VLAN address.

\paragraph{}
In addition, Javid et al. \cite{SDN8} implemented a 2-layer firewall using POX controller. CloudWatcher \cite{SDN3} is a security monitoring framework by which network operators can define a policy to describe a network traffic and describe which security services must be applied to it. Koorevaar et al. \cite{SDN4} proposed an framework for leveraging SDN for automatic security policy enforcement using EEL-tags. These tags are added into the VM's flow by hypervisor. By means of these added EEL tags, they can implement the associated security policy. However, this work heavily relies on trustful hypervisor, thus the portability of method is a big problem need to be considered.

\section{Conclusion}

\paragraph{}
In order to effectively address the security threats of the Internet of things, we propose a SOM-based DDoS defense mechanism using SDN in this paper. To validate the feasibility of the mechanism proposed in this paper, we use POX controller and Mininet emulator to implement an experimental system. We evaluate the SOM-based DDoS defense mechanism. The final experimental results show that the mechanism can effectively detect the DDoS attack sources, and can implement effective dynamic defense after successfully detecting the attack sources, so that it can effectively enhance the security of the Internet of things. However, there exists some limitations in current research work. Because Mininet is difficult to deploy more complex DDoS attack testing toolkits, thus we can only use Ping flood instruction to simulate real DDoS attacks towards the Internet of things. In the future, we intend to deploy the framework in a real network environment, and get more accurate evaluation results.

\section*{Acknowledgments}

\paragraph{}
This paper has been sponsored and supported by National Natural Science Foundation of China (Grant No.61772270), partially supported by National Natural Science Foundation of China (Grant No.61602262).

\section*{References}

\bibliographystyle{elsarticle-num}
\bibliography{test}

\begin{thebibliography}{10}
\expandafter\ifx\csname url\endcsname\relax
  \def\url#1{\texttt{#1}}\fi
\expandafter\ifx\csname urlprefix\endcsname\relax\def\urlprefix{URL }\fi
\expandafter\ifx\csname href\endcsname\relax
  \def\href#1#2{#2} \def\path#1{#1}\fi

\bibitem{iot1}
A.~Rayes, S.~Samer, Internet of things from hype to reality: Internet of things
  security and privacy, Springer International Publishing
  10.1007/978-3-319-44860-2~(Chapter 8) (2017) 195--223.

\bibitem{iot3}
Alibaba, \href{https://yq.aliyun.com/articles/86091}{The security report of
  internet of things (2015)} (2015).
\newline\urlprefix\url{https://yq.aliyun.com/articles/86091}

\bibitem{iot4}
Dyn, \href{https://www.linkedin.com/pulse/dyn-ddos-att
  ack-lessons-learned-small-businesses-bruce-parkman/}{Lessons learned for
  small businesses} (2016).
\newline\urlprefix\url{https://www.linkedin.com/pulse/dyn-ddos-att
  ack-lessons-learned-small-businesses-bruce-parkman/}

\bibitem{SZWang}
S.~Wang, X.~Hu, P.~S. Yu, Z.~Li, Mmrate: inferring multi-aspect diffusion
  networks with multi-pattern cascades., in: KDD, 2014, pp. 1246--1255.

\bibitem{iot2}
L.~Atzori, A.~Iera, G.~Morabito, The internet of things: A survey, Computer
  Networks 54~(15) (2010) 2787--2805.

\bibitem{SDN1}
N.~Mckeown, T.~Anderson, H.~Balakrishnan, G.~M. Parulkar, L.~L. Peterson,
  J.~Rexford, S.~Shenker, J.~S. Turner, Openflow: Enabling innovation in campus
  networks, Acm Sigcomm Computer Communication Review 38~(2) (2008) 69--74.

\bibitem{SOM}
T.~Kohonen, Self-organized formation of topologically correct feature maps, MIT
  Press, 1988.

\bibitem{SDN9}
H.~Hu, W.~Han, G.-J. Ahn, Z.~Zhao, Flowguard: Building robust firewalls for
  software-defined networks, in: ACM SIGCOMM Workshop on Hot Topics in Software
  Defined Networking, 2014.

\bibitem{SDN5}
P.Porras, S.Shin, V.Yegneswaran, M.Fong, M.Tyson, G.Gu, A security enforcement
  kernel for openflow networks, 2012, pp. 121--126.

\bibitem{SDN7}
M.~Suh, S.~H. Park, B.~Lee, S.~Yang, Building firewall over the
  software-defined network controller, in: International Conference on Advanced
  Communication Technology, 2014.

\bibitem{SDN6}
M.~Koerner, O.~Kao, Mac based dynamic vlan tagging with openflow for wlan
  access networks, Procedia Computer Science~(94) (2016) 497--501.

\bibitem{SDN8}
T.~Javid, T.~Riaz, A.~Rasheed, A layer2 firewall for software defined network,
  in: IEEE Information Assurance and Cyber Security, 2014.

\bibitem{SDN3}
S.Shin, G.Gu, Cloudwatcher: Network security monitoring using openflow in
  dynamic cloud networks, in: IEEE International Conference on Network
  Protocols, 2012, pp. 1--6.

\bibitem{SDN4}
T.Koorevaar, Dynamic enforcement of security policies in multi-tenant cloud
  networks, Master's Thesis.

\bibitem{SDN2}
K.~Benzekki, A.~El~Fergougui, A.~Elbelrhiti~Elalaoui, Software-defined
  networking (sdn): a survey, Security and Communication Networks 9~(18) (2016)
  5803--5833.

\end{thebibliography}



%



\end{document}